\documentclass[runningheads]{svmult}

\usepackage{makeidx}   
\usepackage{graphicx}  
\usepackage{subeqnar}  
\usepackage{multicol}  
\usepackage{physprbb}  
\makeindex             

\newcommand{\etal}{{\it et al.\ }}
\newcommand{\lta}{\stackrel{<}{\scriptstyle\sim}}
\newcommand{\gta}{\stackrel{>}{\scriptstyle\sim}}

\begin{document}
\title*{Chemical Evolution and Starbursts}
\toctitle{Chemical Evolution and Starbursts}
%
%
\titlerunning{Chemical Evolution and Starbursts}
%
\author{Uta Fritze -- v. Alvensleben}
\authorrunning{U. Fritze -- v. Alvensleben}
%
%
\institute{Universit\"atssternwarte G\"ottingen, Geismarlandstr. 11, 
37083 G\"ottingen, Germany}

\maketitle              

\begin{abstract}
The first part of this paper deals with the impact of nonsolar and -- for 
late-type, dwarf, and high redshift galaxies -- generally subsolar 
abundances on the interpretation of observational data for starburst galaxies. 
It points out the differences in colors, luminosities, emission lines, etc. 
obtained from a model using low metallicity input physics for a starburst on 
top of the stellar population of a galaxy as compared to an otherwise 
identical model using solar metallicity input physics only. 

The second part deals with the chemical evolution during a starburst and contrasts model predictions with observational clues. 

\end{abstract}

\section{Abundance Effects on Starbursts}
The chemical abundances in the gas of a galaxy at the onset of a starburst 
set the initial abundances for the bulk of the burst stars. Only in long 
bursts, e.g. like those triggered by interactions between masive galaxies, 
may later generations of burst stars incorporate SN\,II products from earlier 
ones. 

Starbursts in giant galaxies, e.g. triggered by an interaction or merger event, are the more spectacular the more gas is available. In the local Universe, a broad anticorrelation is observed between the size and the metallicity of a galaxy's gas reservoir. Hence, the strongest bursts can be excited in gas rich galaxies which genuinely are of low metallicity. Dwarf galaxies as well as young galaxies generally have lower abundances than today's giant galaxies. 

We will show that for a burst of given strength, the spectrophotometric as well as the chemical evolution are significantly dependent on the metallicity. Interpretation of starburst galaxy observations therefore requires comparison with models of appropriate metallicity. 

\subsection{Abundances in Local and High Redshift Galaxies}
While already in the Milky Way, the global average stellar and gas abundances 
both are subsolar ($\sim \frac{1}{2}$ solar), late type galaxies with their 
huge gas reservoirs show significantly lower abundances still
\cite{roc98fva,kil94fva,zar94fva,fer98fva,vzee98fva}. 

So, strong starbursts in local {\bf giant gas-rich galaxies}, e.g. triggered 
by interactions or mergers like in NGC 4038/39 or NGC 7252, are to be 
described by models accounting for the moderately subsolar 
metallicity of the gas in these objects. Burst durations in giant 
galaxies are typically of the order of the dynamical timescale 
${\rm \tau_B \sim t_{dyn} \sim 10^8}$ yr, i.e. long compared to the most 
massive stars' lifetimes. Hence self-enrichment in SN\,II products like 
O, Mg, and other $\alpha$-elements during bursts may be important for 
giant galaxies. Depending on the cooling timescale, these SN\,II products 
may even be incorporated into burst stars that form after the first 
generation of massive burst stars has already died. 

{\bf Dwarf galaxies} in the local Universe, according to the 
luminosity -- metallicity relations established both for the stellar 
metallicities of dwarf elliptical and spheroidal galaxies and for the 
gas metallicities of dwarf irregular galaxies, show significantly subsolar 
abundances that extend down to few percent solar \cite{ric95fva,ski89fva}. 
E.g., Blue Compact Dwarf Galaxies ({\bf BCD}s) typically feature 
${\rm \langle Z \rangle \sim \frac{1}{10}~Z_{\odot}}$, extreme examples 
like IZw18 or SBS 0335-052 reach down to ${\rm \sim \frac{1}{40} Z_{\odot}}$ 
\cite{izo99bfva}. Hence undoubtedly, the interpretation of starbursts 
occuring in dwarf galaxies requires models of appropriately low metallicity. 
In recent years, Tidal Dwarf Galaxies ({\bf TDG}s) have been detected in 
rapidly increasing numbers, forming in the tidal tails of massive interacting 
spirals. Compared to dwarf galaxies of comparable luminosity they show 
enhanced metallicities, typically in the 
range ${\rm(\frac{1}{4} - \frac{1}{2})~Z_{\odot}}$, as a result of being 
formed from stars and pre-enriched gas pulled out from their parent 
galaxies \cite{duc99fva}. 
As for giant galaxies, typical burst durations in dwarf galaxies are of the order of the dynamical timescale, and, hence ${\rm \tau_B \sim 10^6}$ yr for dwarf galaxies. These short burst durations -- not longer than the lifetimes of massive stars -- imply that self-enrichment during a burst will not be important. Mass loss due to SN-driven galactic winds, on the other hand, may be important for dwarf galaxies due to their shallower potential wells as compared to giant galaxies and may significantly affect their chemical evolution. The occurence and strength of galactic winds in dwarf galaxies, however, depend on the poorly constrained mass of their dark matter halos. 

Galaxies in the early Universe, of course, have lower metallicity than their 
local counterparts. Lyman Break Galaxies at ${\rm 3 \lta z \lta 4}$, some 
or many of which are observed in phases of enhanced star formation ({\bf SF}), 
show metallicities in the range ${\rm(0.1 - 1)~Z_{\odot}}$ as derived from 
the restframe UV stellar wind lines of their stellar populations
\cite{low97fva,tra97fva} and from their [O\,III] emission lines 
\cite{tep00fva}. The ample supply of neutral gas in Damped Ly$\alpha$ 
Absorbers at ${\rm 0 \leq z \leq 4.4}$ shows abundances from 
${\rm 10^{-3}~Z_{\odot}}$ to ${\rm \lta Z_{\odot}}$ 
\cite{pet94fva,pet99fva}. 

\subsection{Chemical and Spectrophotometric Evolution Models}
In the following we use evolutionary synthesis models to describe the 
spectrophotometric and chemical evolution of starbursts in various types 
of galaxies. The models for undisturbed galaxies of various types are 
parametrised by the respective appropriate star formation histories 
${\rm \Psi(t)}$ and constrained by the requirement to provide agreement 
after $\sim 12$ Gyr of evolution with average colors, luminosities (U ... K), 
emission and absorption line strengths, characteristic H\,II region abundances 
(= measured at ${\rm R_{eff}}$), and gas content of local samples of the 
respective types and with template spectra. Then a burst of given strength 
${\rm b := \frac{\Delta S}{S}}$ (with S : stellar mass at the onset of the 
burst, ${\rm \Delta S}$ : stellar mass added in the burst) and duration 
${\rm \tau_B}$ is assumed to start at some time ${\rm t_B}$. Using sets of 
input physics (stellar evolutionary tracks, lifetimes, stellar yields, 
model atmosphere spectra, and absorption index calibrations) for a range 
of metallicities from $10^{-4}$ to 0.05, models follow the spectral 
evolution from UV -- NIR, the chemical evolution of individual gas phase 
element abundances, as well as the metallicity distribution in the stellar 
population before, during, and after the burst. Models account for the 
finite lifetimes of the stars before they give back enriched material, 
include the contributions of type Ia SNe as described by \cite{mat85afva}
 but they are otherwise kept as simple as possible in order to minimize 
the number of parameters. In particular, they are closed box 1-zone models 
and assume instantaneous and perfect mixing of the recycled gas. 

\subsection{Abundance Effects on Starbursts}
At subsolar metallicities, bursts of a given strength in the same type of 
galaxy lead to higher peak luminosities, bluer optical colors, slower and 
weaker fading and reddening after the burst (Fig.~1), and higher mass 
loss from the burst population (due to shorter stellar lifetimes) as 
compared to models using solar metallicity input physics
\cite{krue94fva,kur99fva,schu01fva}. 
Furthermore, at lower metallicities, spectra look different, emission 
line ratios change \cite{izo99afva}, the emission contribution of the gas 
is higher to UBVR fluxes and lower in JHK bands (Fig.~2), UV stellar wind 
lines are weaker \cite{lei99fva}, and dust extinction less important. 

Hence, in a low metallicity galaxy, the same blue optical colors imply a weaker or older burst than they would in a solar metallicity galaxy.

\begin{figure}[b]
\begin{center}
\includegraphics[width=.5\textwidth]{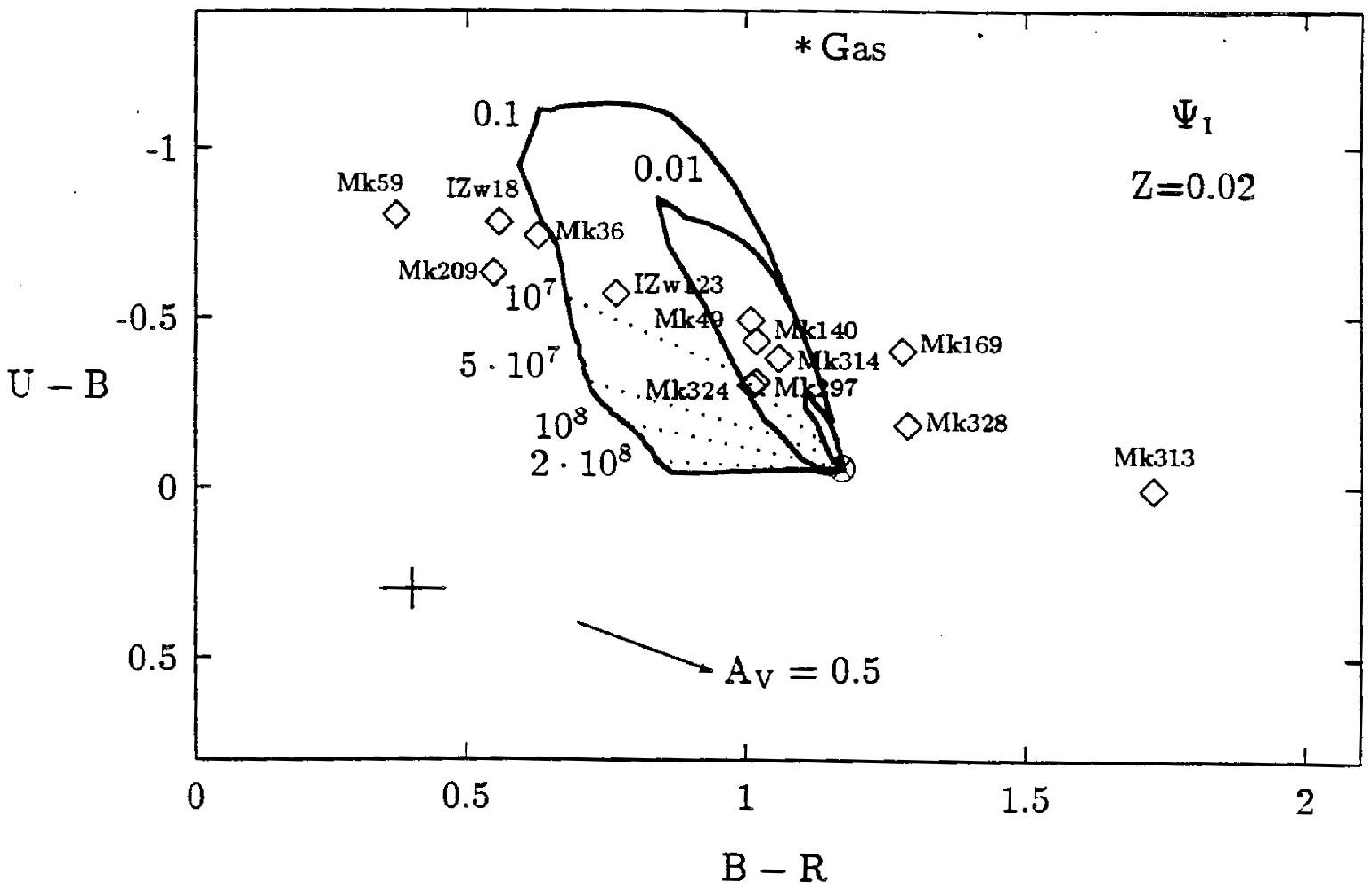}\includegraphics[width=.5\textwidth]{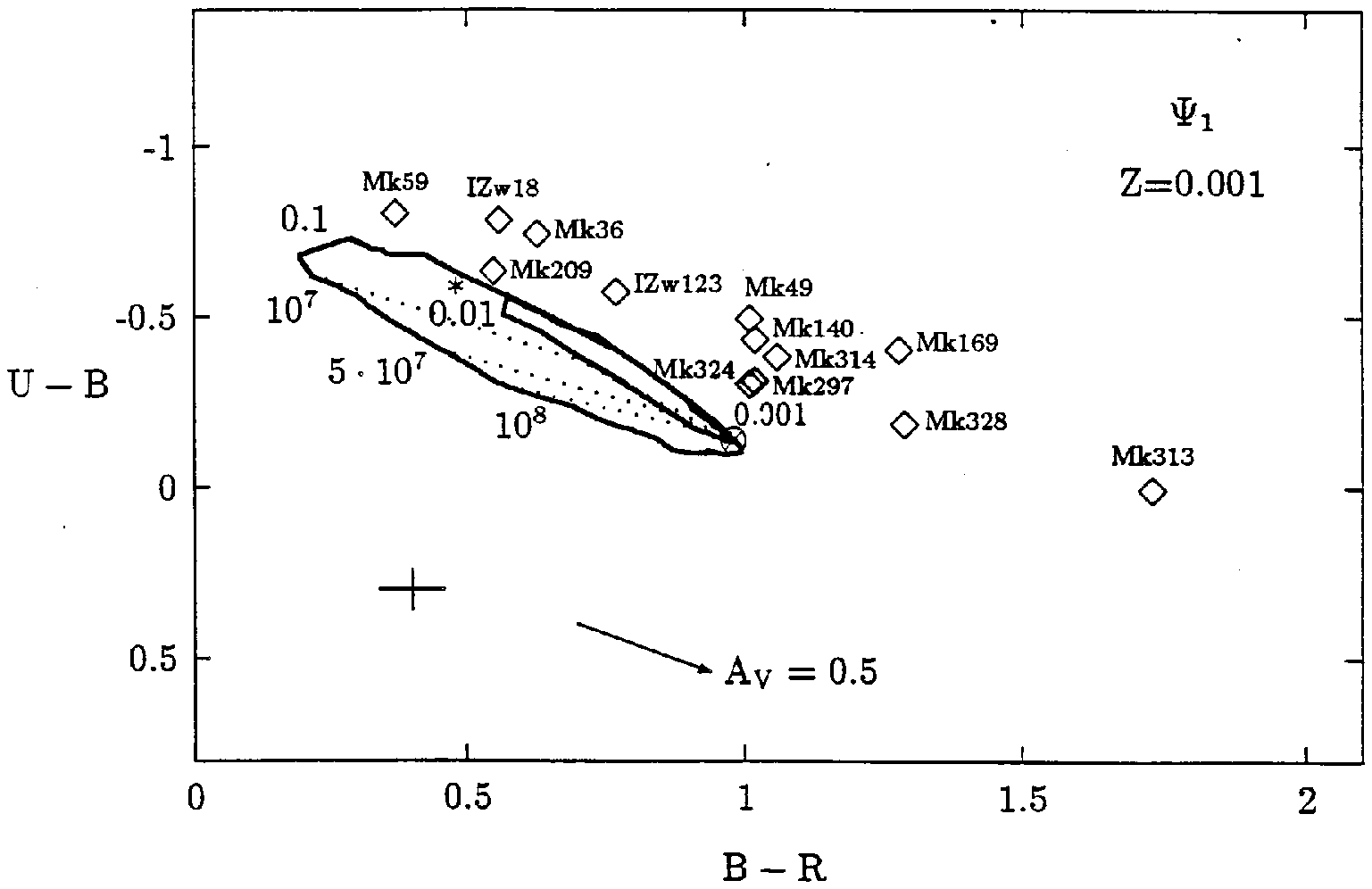}
\end{center}
\caption[]{Color-color diagram for starbust models at two different metallicities and 3 burst strengths each (b=0.1, 0.01, 0.001) on top of a galaxy with const. SF rate. $\bigotimes$ marks the galaxy before the burst, $\ast$ the color of the pure gas spectrum. BCD data are from \cite{thu83fva}.}
\label{eps1fva}
\end{figure}

\begin{figure}[t]
\begin{center}
\includegraphics[angle=270,width=.5\textwidth]{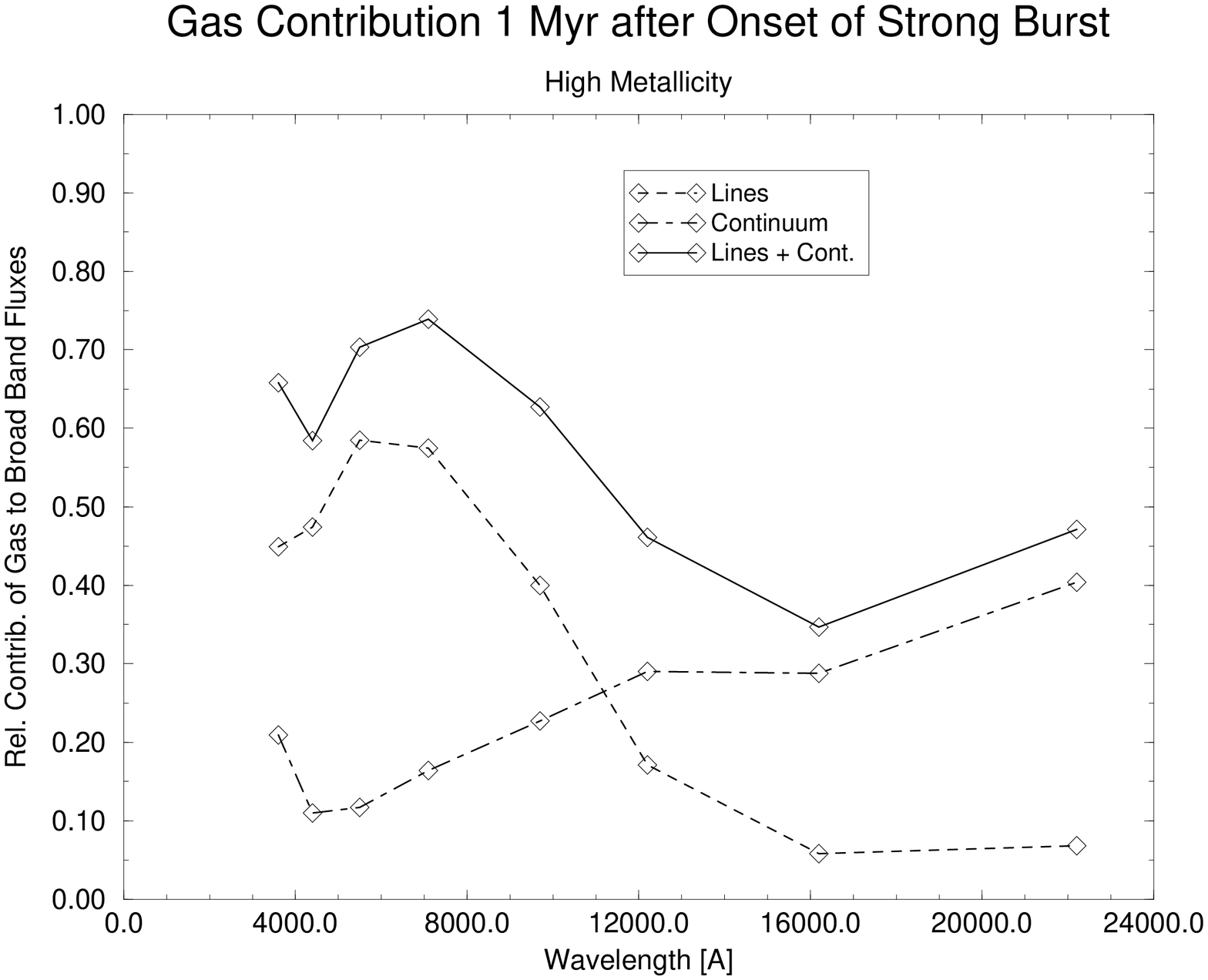}\includegraphics[angle=270,width=.5\textwidth]{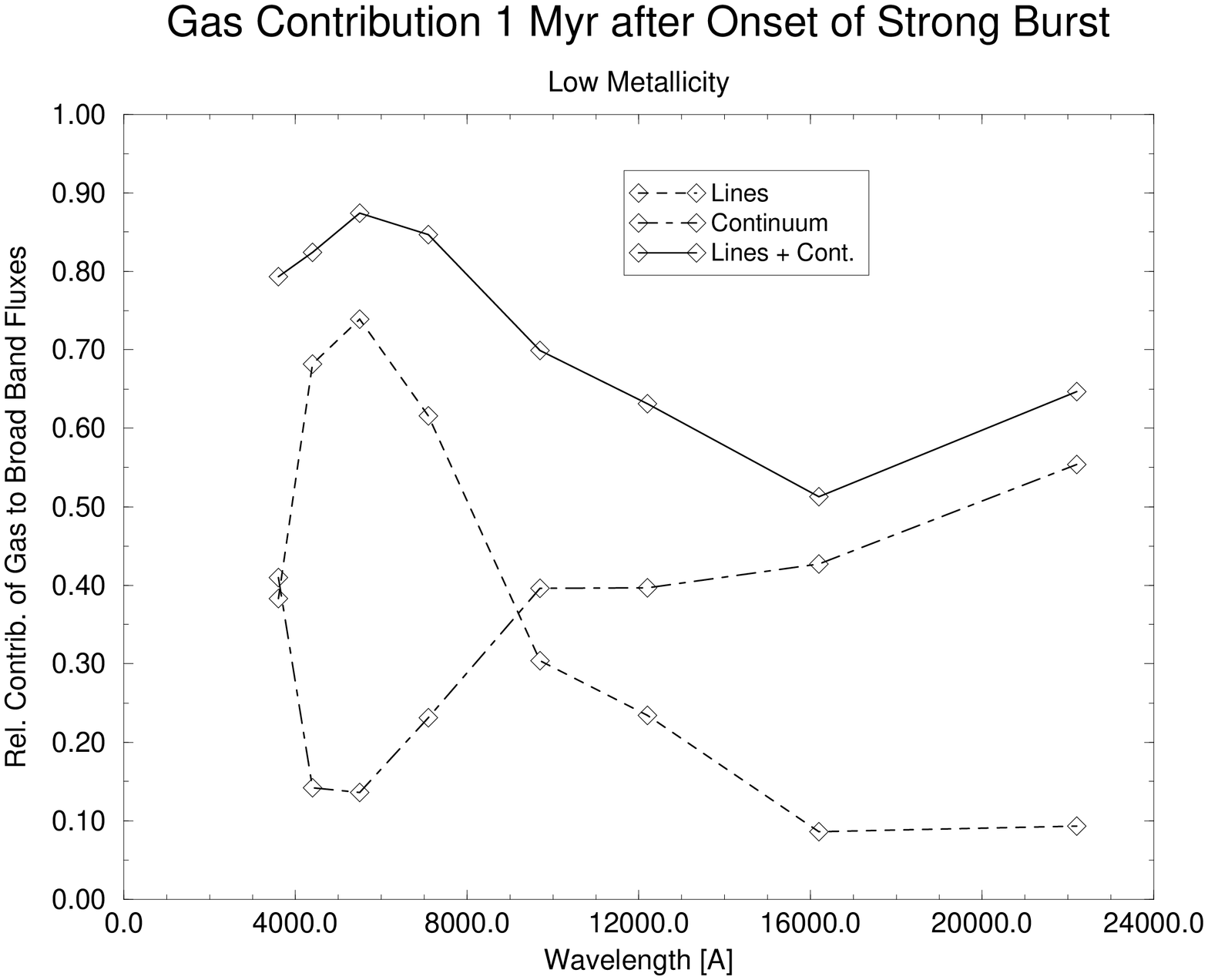}
\end{center}
\caption[]{Relative contribution of the gaseous emission to broad band fluxes UBVRIJHK 1 Myr after the onset of a strong burst (b=0.4). Left for solar metallicity, right for Z=0.001.}
\label{eps2fva}
\end{figure}

\section{Chemical Evolution in a Starburst}

\subsection{Model Predictions}
The chemical evolution -- both in terms of ISM enrichment and metallicities 
of stars formed in the burst -- depends on the burst strength, the size of 
the gas reservoir, and on the (time of) occurence or non-occurence of 
SN-driven galactic winds or superwinds. The latter, unfortunately, depends 
on several poorly known conditions, e.g. on the masses of DM halos, the 
geometry of the starburst region, etc. Note that the same number of burst 
stars can cause vastly different increases of the ISM metallicity depending 
on whether they shed their enrichment products into a small or a large gas 
reservoir. During a burst, individual element abundances in the ISM increase 
on individual timescales, depending on the nucleosynthetic origin of the 
respective element. Oxygen and other typical SN\,II products 
($\alpha$-elements) are restored on very short timescales given by the 
lifetimes of massive stars. Thus the abundances of oxygen and other 
$\alpha$-elements start increasing shortly after the onset of the burst 
and do not continue for long after its end. Elements that are predominantly 
synthesized in intermediate mass stars, like C or N, as well as elements 
that have important contributions from type Ia SNe, such as Fe, only start 
enriching with certain time delays after the beginning of the burst and 
continue to do so for up to some Gyr after the end of the burst. This effect 
leads to strong changes in the element ratios between elements of different 
nucleosynthetic origin. As e.g. \cite{mat85bfva} showed, the large 
range of N/O-ratios at essentially all oxygen abundances among dwarf galaxies, 
Galactic and extragalactic HII regions may be explained by successive bursts 
of SF. During a short burst, the oxygen abundance increases and, hence N/O 
decreases while nitrogen is not changed yet. The nitrogen abundance only 
starts increasing after the end of the burst, leading to an increase of N/O 
at essentially constant oxygen abundance during up to a few Gyr. 

A detailed understanding of the chemical evolution in the course of a 
starburst not only requires a consistent chemodynamical model but also 
the consideration of the multi-phase nature of the ISM, including a full 
description of all kinds of transition processes between the hot X-ray gas, 
the warm and neutral (HI) components and the cold and clumpy molecular gas. 
This latter component, often traced by CO, is increasingly difficult to 
observe in low metallicity dwarf galaxies. A consistent and complete model 
including all these phases and processes is still to be developed 
\cite{rie00fva}.

\subsection{Observational Clues}
\subsubsection{ISM Abundances}

H\,II region abundances in BCDs are used to obtain the metallicity of the 
stars formed in the burst. Burst strengths in BCDs are weak as compared to 
those in massive gas rich interacting galaxies, increasing the stellar mass 
by typically few percent or less \cite{krue95fva}. H\,I reservoirs, on the 
other hand, are large in BCDs. So, the metallicity increase in the burst and, 
hence, the metallicity difference between preburst and burst stars both are 
relatively small. This explains why models using input physics for the H\,II 
region metallicity observed in a BCD generally also allow for good agreement 
with its spectral energy distribution which, at short wavelengths, is 
dominated by burst stars while at long wavelengths (JHK) it is mostly due 
to the preburst component \cite{krue95fva}. 

TDGs, on the other hand, with their observed H\,II region abundances enhanced 
over those of dwarf galaxies of comparable luminosity, in most cases turn 
out to be well describable with models including an old stellar population 
from a late type galaxy plus a burst star component of $\lta \frac{1}{2}$ 
solar metallicity \cite{wei00fva}. Their H\,II region abundances indeed 
are ${\rm [O/H]_{TDG} \gta [O/H]_{spiral}^{ISM}}$. 

\subsubsection{Stellar and Star Cluster Abundances}

Stellar abundances for starburst populations are difficult to disentangle observationally from those of the preburst stars. Star clusters formed in starbursts in some dwarf galaxies and -- in huge numbers -- in the strong bursts accompanying gas rich galaxy mergers, however, provide a powerful tool to study the chemical enrichment process in these bursts. As opposed to field stars, they well separate from the preburst population and can be analysed individually. 

Models predict their abundances on the basis of the progenitor galaxy ISM 
abundances and gas reservoirs, and of the strength and duration of the burst. 
E.g. for the young and very young star cluster populations in NGC 7252 and 
NGC 4038/39 average metallicties of ${\rm (\frac{1}{2} ~-~ 1) Z_{\odot}}$ 
were predicted as well as some $\alpha$-enhancement for the youngest clusters 
in the almost 1 Gyr old burst in NGC 7252 \cite{fva94afva,fva95fva}. 
\cite{kur99fva} and \cite{fva00fva} show how significant the metallicity of 
a star cluster is in interpreting its photometric data, e.g. for age-dating 
and mass estimates. Young star clusters are bright -- at ages of $\sim 10^8$ 
yr typically 4 mag brighter than old globular clusters ({\bf GC}s) of the 
same mass. 10 m telescope spectroscopy thus directly gives access to 
metallicities, ages, velocity dispersions for kinematic mass estimates, ... 
\cite{ho96fva,schw93fva,schw98fva}.

For clusters older than few $10^8$ yr which no longer feature emission lines, 
stellar absorption features in comparison with theoretical model calibrations 
at the appropriate young ages \cite{kur99fva,schu01fva} give information 
about individual element abundances and abundance ratios. With spectroscopy 
of reasonable samples the metallicity and age distributions among young star 
clusters and their spatial variations will give information about the 
dynamics of the burst and its detailed enrichment process. The limiting 
factor for this kind of observations is the strong and spatially variable 
galaxy background. 

MOS of old GCs is within reach of 10m telescopes out to Virgo cluster 
distances. The metallicity distributions this will reveal for the old GC 
populations of elliptical and S0 galaxies will teach us a lot about their 
formation processes \cite{zep93fva,geb99fva}.

\section{Conclusions}
Accounting for the appropriate metallicities in evolutionary synthesis models is essential for the interpretation of starburst galaxies as is the consideration of dust (see S. Charlot, {\sl this vol.}). 

In a burst, we see to first order star formation and ISM abundances at the metallicity of the gas in the preburst galaxy. Most clearly this is seen on star clusters rather than on field stars. Only to second order metallicity and, in particular, $\alpha$-enhancements are expected in the long lasting bursts in massive interacting gas rich galaxies. Again these are best studied on individual star clusters.

\bigskip\noindent
{\small \bf Acknowledgement.} {\small I gratefully acknowledge partial financial support from the organisers.}

\end{document}